\documentclass[usenatbib]{mn2e}

\usepackage[dvips]{graphicx}
\usepackage{amssymb}
\usepackage{xcolor}    
\usepackage[normalem]{ulem}

\def\etal{{\rm et al. }}

\def\kpc{{h^{-1} \rm kpc}}
\def\pc{{h^{-1} \rm pc}}
\def\kms{{\rm km\, s^{-1}}}
\newcommand\aap{{\em A}\&{\em A}}
\newcommand\aaps{{\em A}\&{\em AS}}
\newcommand\aj{{\em AJ}}
\newcommand\apj{{\em ApJ}}

\newcommand\apjs{{\em ApJS}}

\newcommand\mn{{\em MNRAS}}

\newcommand\pasp{{\em PASP}}

\begin{document}

\title{LINER galaxy properties and the local environment}

\author[Coldwell \etal]{Georgina V. Coldwell$^{1}$, Sol Alonso$^{1}$, Fernanda Duplancic$^{1}$ and Valeria Mesa$^{2}$\\
$^{1}$ Departamento de Geof\'{i}sica y Astronom\'{i}a, CONICET, Facultad de Ciencias Exactas, F\'{i}sicas y Naturales, Universidad Nacional \\
de San Juan, Av. Ignacio de la Roza 590 (O), J5402DCS, Rivadavia, San Juan, Argentina\\
$^{2}$Instituto Argentino de Nivolog\'{i}a, Glaciolog\'{i}a y Ciencias Ambientales (IANIGLA-CCT Mendoza, CONICET), Parque Gral San Mart\'{i}n, \\
CC 330, CP 5500, Mendoza, Argentina}

\date{\today}

\pagerange{\pageref{firstpage}--\pageref{lastpage}}

\maketitle

\label{firstpage}

\begin{abstract}

We analyse the properties of a sample of 5560 LINER galaxies selected from SDSS-DR12, at low redshift, for a complete range of local density environments. The host LINER galaxies were studied and compared with a well-defined control sample of 5553 non-LINER galaxies matched in redshift, luminosity, morphology and local density. By studying the distributions of galaxy colours and stellar age population we find that LINERs are redder and older than the control sample in the wide range of densities. In addition, LINERs are older than the control sample, at a given galaxy colour, indicating that some external process could have accelerated the evolution of the stellar population.

The analysis of the host properties shows that the control sample exhibits a strong relation between the colours, ages and the local density, while more than 90\% of the LINERs are redder and older than the mean values, independently of the neighbourhood density. Furthermore, a detailed study in three local density ranges shows that, while control sample galaxies are redder and older as a function of its stellar mass and density, LINER galaxies mismatch the known morphology-density relation shown by galaxies without low-ionization features.
The results give support to the contribution of hot and old stars as viable  mechanisms for the low ionization emission although the presence of nuclear activity is not discarded.

\end{abstract}


\section{Introduction}
\label{intro}

Galaxies with low-ionization nuclear emission-line regions 
(LINER) were firstly described by \cite{Heck80} as a class of extragalactic objects 
with optical spectra dominated by enhanced low-ionization OI($\lambda 6300$) and 
NII($\lambda 6548,6583$) lines.
Therefore, LINERs were defined by intensity ratios of optical emission lines, namely: 
(1) I([O II] $\lambda$ 3727)/I([O III] $\lambda$ 5007) $\geq$ 1 where [O II]
$\lambda$ 3727 is  used to designate the [O II] $\lambda$ $\lambda$ 3726, 3729
doublet, and (2) I([O I] $\lambda$ 6300) /I([O III] $\lambda$ 5007) $\geq$ 1/3.
These low-ionization features were found in about 30 \% of the nearby galaxies \citep{Heck80, HoFilSar97}, reaching 50\% in elliptical 
passive galaxies \citep{Fil86,Goud94,Yan06,CapB11}.


Different mechanisms were proposed to explain the nature of LINER galaxies. 
Their low-ionization emission lines could be powered by 
(1)shock-heated gas \citep{Heck80,DopSut95}, which is the less likely 
scenario due the fact that gas velocity dispersion commonly falls 
below the value required to explain the observed level of spectral ionization 
\citep{Ho03}; (2) photo-ionization by an active central black hole \citep{Groves04}, such as an Active Galactic Nuclei (AGN); and (3) stellar photo-ionization by hot 'O' stars \citep{FilTer92} 
or by old post-asymptotic giant branch (post-AGB) stars \citep{Bin94,Stas08}.

Several works agree with the assertion that AGN photoionization is the dominant mechanism for LINER emission. In this line,
\cite{Kewley06} establishes that LINERs seem to populate the low luminosity end of the AGN distribution, 
where radiatively inefficient accretion flows and external obscuring matter \citep{GM09b,dudik09} 
may cause the optical extinction. Thus, LINERs are less luminous than Seyfert galaxies and share 
similar spectral characteristics, with the remarkable exception that LINERs show 
enhanced low-ionization OI($\lambda 6300$) and NII($\lambda 6548,6583$) lines 
\citep{Heck80}. Accordingly, observations at radio \citep{nagar05} and X-ray wavelengths 
\citep{GM09a} provide strong support for an AGN as the origin of LINER emission.

Regarding the stellar photo-ionization hypothesis, several authors suggest that hot post-AGB stars and white dwarfs could provide enough 
ionization to explain the LINER emission \citep{Bin94,sodre99,Stas08}. Moreover, these stellar ionization sources are located in spatially extended regions around the nucleus as the $H\alpha$ and $H\beta$ brightness profiles do not decrease with $r^{-2}$,
as could be expected for the radial dependency of a central ionization source such as an AGN \citep{sarzi08,YanB12}. 

Recently, the surveys MANGA (Mapping Nearby Galaxies at Apache Point Observatory, \cite{bundy15}) and CALIFA (The Calar Alto Legacy Integral Field Area, \cite{sanchez12}) allowed to obtain valuable spatially resolved observations giving light to the issues related to the LINER galaxies. By using MANGA \cite{belfiore16} studied the spatially resolved excitation properties of the ionized gas in a sample of 646 galaxies demonstrating the presence of extended low ionization emission-line regions on kpc scales. Thus, they introduce the acronym LIER, instead of LINER, to highlight that low ionization lines in most of the galaxies are not nuclear, being post-AGB stars the most likely candidates for their emission.

Furthermore, \cite{sin13} analysed the radial emission-line surface brightness profiles of a sample of LINERs, selected from CALIFA survey, in comparison to the expected profile for AGNs. Their finding were consistent with emission corresponding to extended power sources, in agreement with the work of \cite{papa13}, showing strong evidence in favour of the post-AGB stars hypothesis.
Based on the results of the cited works it seems to be very useful to properly distinguish between ``nuclear'' and ``extended'' LINERs, although it doesn't seem to be trivial. \cite{YanB12} consider the scale over which the spectrum is taken, and define nuclear LINERs at scales lower than $200 \pc$ and extended sources when considering scales grater than $1 \kpc$. In fact, they argued that this distinction strongly depends on the physical aperture of the observation instrument. For SDSS galaxies, with $z\sim0.1$ most LINERs are spatially extended. Also, the authors state that the host galaxies of nuclear line-emitting regions and those of extended line-emitting regions are largely the same population.

The possibility that several different mechanisms power simultaneously the low-ionization emission lines is also considered.
For example, HST observations have found that both unresolved nuclear and extended H$\alpha$ emission are present in the majority of nearby LINERs \citep{masegosa}.
Moreover, \cite{graves07} found that red-sequence galaxies with LINER emission 
are younger than their quiescent counterparts, suggesting a connection between the star formation history and the mechanism generating the low ionization emission. Nevertheless, there is no certainty of the predominant mechanism in LINER galaxies.
In addition \cite{erac10} observed a sample of 35 LINERs, on small scales of order $\sim 200 \pc$ from the nucleus, and found  that photoionization by post-AGB stars is an important power source, since in more than half of the LINERs it provides more ionizing photons than the AGN and enough ionizing photons to power the emission lines in one third of them. Besides, photoionization from the AGN power the gas emission on scales of few tens of parsec although it is not enough efficient on larger scales suggesting that different processes could dominate on different scales.

To unravel the nature of the LINER emission the study of environmental factors could provide important clues, since numerous galaxy properties correlate with the density 
environment.
In this way, several studies  \citep{PB06,coldwell09,padilla10,coldwell14} have found that AGN  environments do not seem to follow the morphology-density relation proposed by \cite{Dress80}.
Then, if the photoionization from AGNs is one of the mechanisms for the LINER emission we could expect that, at least, LINERs follow a similar trend that Seyfert 2 galaxies.
On this way,  \cite{coldwell17} analysed the occurrence of SDSS-DR7 LINER galaxies with respect to their proximity to galaxy groups. The main results have shown that a higher percentage of LINER can be found populating low-density environment. Besides, if LINERs belong to high virial mass galaxy groups, they do not follow the expected morphology-density relation.

In this paper we study the behaviour of LINER galaxy properties compared with a well defined control sample selected
to match the density environment, redshift, luminosity and morphology.
The layout of this paper is as follows: in Section 2 we 
briefly describe the data selection, the LINER classification scheme and the procedure used to 
construct the control sample. In section 3, the analysis of galaxy properties and their morphology density relation 
is presented. Finally, in Section 4 we discuss the results and draw our 
conclusions. Throughout this paper, we have assumed a $\Lambda$-dominated 
cosmology, with $\Omega_{m} = 0.3$, $\Omega_{\lambda} = 0.7$ and 
$H_0 = 100 \kms Mpc^{-1}$.

\begin{figure*}
\includegraphics[height=60mm,width=160mm]{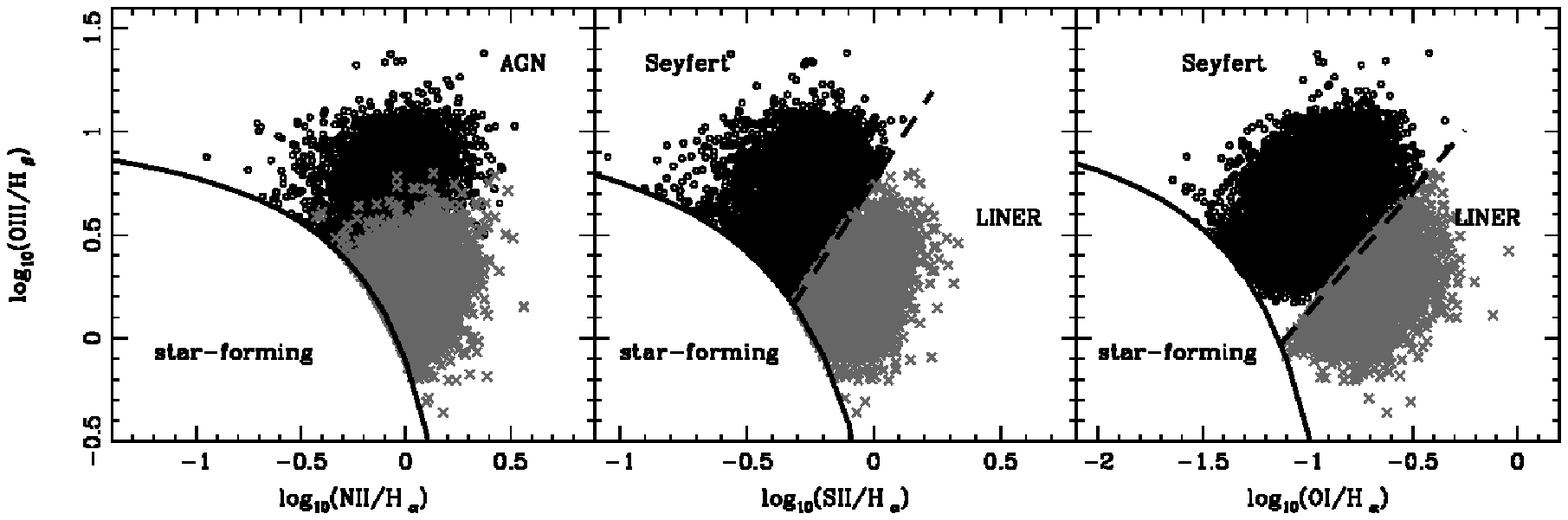}
\caption{BPT diagrams of the selection criteria defined by Kewley \etal (2006) 
used to classify emission-line galaxies as Seyfert or LINER.
The panels show the line ratios $\log([\rm OIII]/\rm H\beta)$ vs $\log(\rm [NII/H\alpha])$ (left),
$\log([\rm OIII]/\rm H\beta)$ vs $\log(\rm [SII/H\alpha]$ (central) and 
$\log([\rm OIII]/\rm H\beta)$ vs $\log(\rm [OI/H\alpha]$ (right).
Seyferts are indicated by black dots and LINER galaxies by grey crosses. The solid lines separate
star-forming galaxies from AGN and the dashed lines represent the Seyfert-LINER demarcation.}
\label{fig1}
\end{figure*}

\section{Data and Sample Selection}

The sample of galaxies used in this work were drawn from the Data Release 12 of Sloan Digital 
Sky Survey\footnote{https://www.sdss3.org/dr12/} \citep[SDSS-DR12,][]{Alam15}. This survey 
covers 14555 square degrees of sky and includes imaging in 5 broad bands ($ugriz$), reduced and calibrated 
using the final set of SDSS pipelines. The SDSS-DR12 provides spectroscopy of roughly more than two million 
galaxies including additional galaxy and quasar spectra from the SDSS-III Baryon Oscillation Spectroscopic 
Survey \citep[BOSS,][]{Dawson13}. 

All catalogue data were obtained through $\rm SQL$ queries in $\rm CasJobs$\footnote{http://skyserver.sdss.org/casjobs/}.
We select galaxies with spectroscopic information and extinction corrected model magnitudes which are more appropriated for extended objects (e.g. galaxies) and also provide more robust galaxy colours. The magnitudes are k-corrected using the empirical k-corrections 
presented by \cite{Omill11}. 

The physical properties of the galaxies used in this analysis were taken from the MPA-JHU\footnote{http://www.sdss.org/dr12/spectro/galaxy\_mpajhu/}.
The procedures to estimate parameters such as stellar masses, emission-line fluxes, stellar age indicators, etc. 
are based on the methods of \cite{brinch04}, \cite{tremonti04} and \cite{kauff03b}.

\subsection{LINER Sample}
\label{sec:sel}

The LINER selection was performed by using the publicly available emission-line fluxes corrected for optical reddening using the Balmer decrement and the \cite{calzetti00} dust curve. We assume an $R_V=A_V/E(B-V)=3.1$ and an intrinsic Balmer decrement $(H\alpha/H\beta)_{0}=3.1$ \citep{OM89}. 
 Details about the line measurements are described by \cite{tremonti04} and \cite{brinch04}.

To select LINERs from the emission-line galaxy sample we only include galaxies with  $S/N > 2$ 
for all the lines involved in the current analysis. This conservative criteria only moderately reduces the size of the sample
by assuring a more reliable selection of LINER objects. 
Following \cite{coldwell17}  the sample was restricted to have a redshift range of $0.04 < z < 0.1$ and we distinguished Seyfert 2, LINER and star-forming galaxies by using the three standard \cite[][: BPT]{BPT81} line-ratio diagrams. Thus, we adopted  the empirical demarcation lines from \cite{Kewley01,Kewley06} for the AGN/starburst and for the LINER/Seyfert separations.

Finally, we obtained an effective sample of 5581 LINER objects. It is important to emphasize that the spectral resolution of the circular fibre used to measure SDSS-DR12 spectra, at the mean redshift of this sample, allows us to select mainly extended LINERs, although ``nuclear'' LINERs could be hidden in our sample.
The discriminated samples and the selection criteria are shown in the three BPT diagrams of Figure \ref{fig1}.

\subsection{Control Sample}

In a recent work \cite{coldwell17} analysed a sample of SDSS-DR7 LINER galaxies with respect to its proximity to galaxy groups taking into account that the physical properties of LINER host galaxies have morphologies, colours and ages corresponding to galaxies residing in the central regions of galaxy groups or clusters.
In this study the authors used a well-defined control sample to obtain strong 
conclusions respect to the relation between the low-ionization features of LINER galaxies and the high density environment. However, the results show that LINERs are more likely to populate low-density environments 
in spite of their morphology, by comparing with the neighbourhood of galaxies with identical morphological features but not LINER emission.

From a different perspective, in this paper we explore the properties of LINER galaxies, taken from SDSS-DR12, 
in a complete range of densities. Furthermore, with the aim of revealing the relation between the low-ionization emission 
line mechanisms and the environment, a suitable control sample of galaxies is used for comparison. In this sense 
\cite{Perez09}, by using SDSS mock galaxy catalogues built from the Millennium Simulation,  showed 
that a suitable control sample for galaxies in pairs should be selected (at least) 
with matched distributions of redshift, morphology, stellar mass, and local 
density environment.

\begin{figure}
\includegraphics[height=93mm,width=87mm]{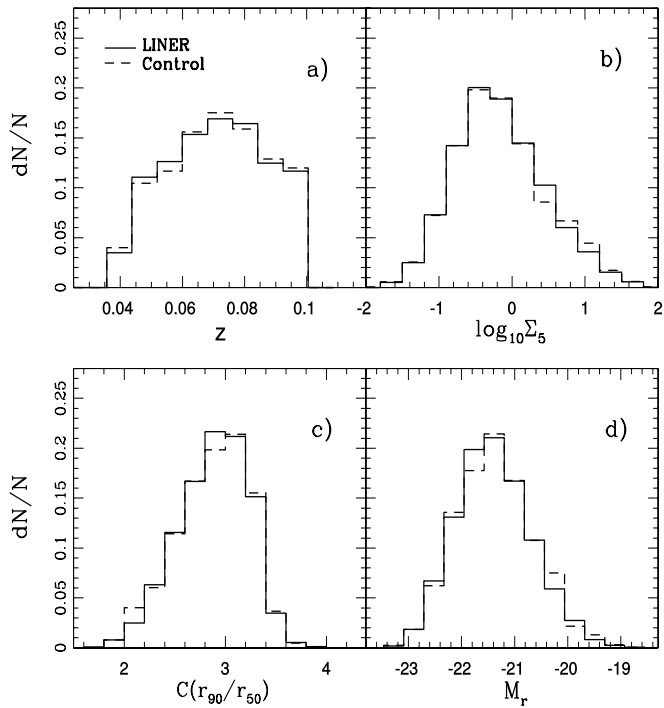}
\caption{Normalized distributions of galaxy properties, and local density, for LINERs (solid line) and control sample (dashed line)}
\label{fig2}
\end{figure}

\begin{figure*}
\includegraphics[height=100mm,width=170mm]{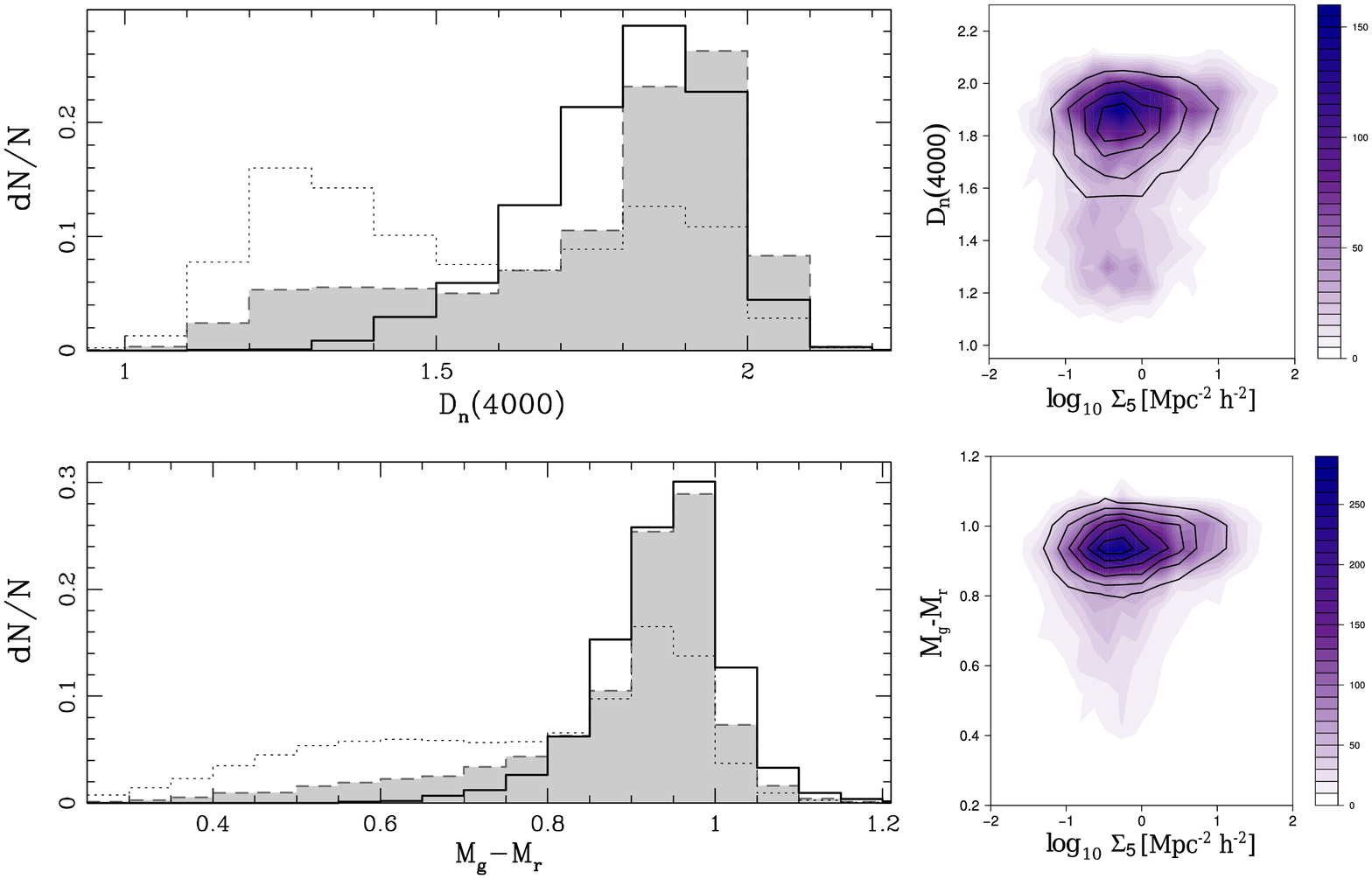}
\caption{Left panel: normalized distributions of stellar age population parameter, $D_n(4000)$, (top) and  
colours, $M_g-M_r$, (bottom) for LINERs (solid line) and control sample (shaded histogram). Dotted line corresponds 
to the MGS in the redshift range of the targets. Right panel:  $D_n(4000)$ versus  
local density, $\Sigma_5$, (top) and $M_g-M_r$ versus $\Sigma_5$ (bottom) for LINERs (solid contours) and control 
sample (coloured density map)}
\label{fig3}
\end{figure*}

Bearing this in mind, we applied
a similar criteria to construct our control sample by selecting
galaxies without low-ionization emission features from SDSS-DR12 with matched distributions of i) redshift, ii) luminosity given for the absolute magnitude in r\_band, iii) concentration index C defined 
as the ratio of the radii containing 90\% and 50\% of the Petrosian flux, respectively 
and iv) the local density, $\Sigma_5$, defined as

\begin{equation}
\Sigma_5=\frac{5}{\pi d_{5}^2},
\end{equation}

where $d_5$ corresponds to the projected distance of the fifth neighbour brighter than $M_r < -20.5$. This 
two-dimensional density estimator use the redshift information to reduce the projection effects and
is useful to characterize the local galaxy density. The advantage of this method is to use a systematically
larger scale in lower-density regions which improves sensitivity and precision at
low densities. We choose a fixed velocity interval of $\Delta V=1000\kms$ to compute
the local density which correspond to galaxies within $\sim 3\sigma$ from the centre
of a galaxy cluster \citep{Balogh04} and this allows the inclusion of galaxies in system
with large velocity dispersion.

In addition, the concentration index $C$ can be correlated with the morphological characteristic of galaxies 
\citep{shima01,strate01}. Thus, galaxies with a de-Vaucouleurs profile have a value of $C \sim 3.3$ and disk galaxies 
have a concentration index $C \sim 2.4$. In Fig. 2 it is shown the normalized distributions of redshift, luminosity, morphology and local density for both, LINER and control samples. The carefully selected control samlpe comprises 5553 galaxies. By matching these parameters we are able to detect, in the free remaining parameters, distinctive features from LINERs likely associated to the low ionization emission mechanisms.

\begin{figure}
\includegraphics[height=85mm,width=85mm]{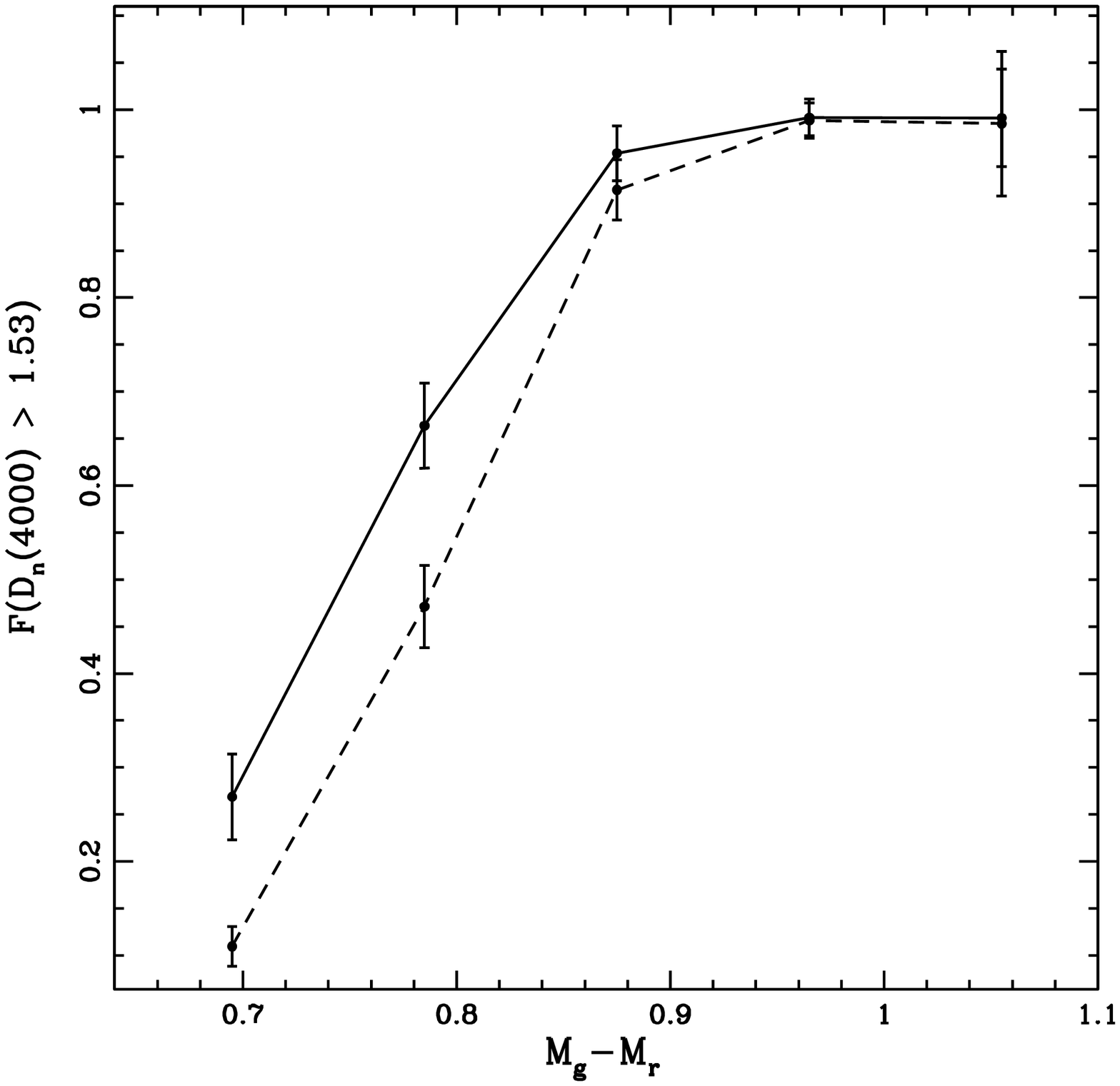}
\caption{Fraction of old ($D_n(4000)> 1.53$) galaxies
as function of the colours, $M_g-M_r$. The solid line corresponds to
the LINER sample and the dashed line to the control sample, respectively.}
\label{fig2B}
\end{figure}

\section{Analysis}

Different studies demonstrated that the host galaxies of LINERs have a bulge-type morphology being more massive and luminous than a general galaxy sample \citep{Yan06,Kewley06,CapB11,coldwell17}.
This early-type galaxies are commonly found in dense galaxy environment as it is suggested by the morphology-density relation
\citep{Dress80,DML01}. However, \cite{coldwell17} have shown that the probability to find LINER galaxies in rich galaxy groups is lower than that found in low-density environment. Thus, LINERs do not seem to follow the expected morphology-density relation.    

From a wide perspective, in this work we use a complete range of densities to explore the dependence of the host galaxy properties with the low-ionization features and the environment since only LINERs close to galaxy groups were analysed by \cite{coldwell17}. Moreover 
exhaustive statistical studies of the LINERs environmental dependence have not been done until now.  

\subsection{Colours and stellar age populations of host galaxies}

Galaxy colours can be used as estimators of the galaxy evolution. So, in clusters the large fraction of red galaxies
indicates an old population of galaxies with a low star formation rate. These evolved sources, as consequence of dynamical processes such as interactions, ram-pressure striping and strangulation \citep{GunnGott1972,Larson1980,Balogh2000} are typically elliptical galaxies. On the contrary, galaxies in poor groups or in the field are bluer and with stronger star formation rate, indicating young sources.

In addition, the break index $\rm D_n(4000)$ \citep{kauff02}, defined as the ratio of the average flux density in the narrow continuum bands ($3850-3950$ and $4000-4100$ \AA), is 
suitably correlated to the mean age of the stellar population in a galaxy. It can be 
used to estimate the star formation rate \citep{brinch04}, where the majority of star formation takes place 
preferentially in galaxies with low $\rm D_n(4000)$ values.

In this section we analyse in detail colours and ages indicators of galaxies in the samples. In the left panels of Figure 3 it is possible to observe the distributions of stellar age populations and colours, given by $\rm D_n(4000)$ 
and $M_g-M_r$ respectively, of LINER and control galaxy samples. Also we plot, the distributions of these parameter for galaxies in the SDSS Main Galaxy Sample (MGS) in the redshift range $0.04 < z < 0.1$. Both samples, LINER and control, are redder and older than MGS galaxies, however a significant difference between them is detected. 
 An evident lack of young and blue galaxies with LINER emission can be observed in both distributions in comparison with the control sample.

To quantify this effect we calculated the fractions of galaxies with stellar age populations $ D_n (4000)> 1.53 $ and colours $ M_g-M_r> 0.76 $. These constraints correspond to the mean values of these parameters for the MGS in the redshift range considered in this work.
Hence, we estimate that 94 \% of LINER galaxies have older stellar populations and redder colours than 
the average, while this percentage decreases up to 78 \% for the control sample galaxies.

In addition, the right panels of Figure 3 show that LINERs are older and have redder colours than the corresponding control 
sample, independently of their local density. Instead of that, galaxies in the control sample have a larger spread of colours 
and ages of their stellar population, being bluer and younger at lower densities.

Furthermore, colours and stellar age show clearly that LINERs belong to a more evolved galaxy population for several ranges of local galaxy density, although the appearance of Fig. 3 suggests the presence of a difference between these two parameters for both samples.
To analyse this effect we quantify any excess of old stellar age population with respect to their galaxy colours by calculating the fraction of galaxies with $D_n (4000)> 1.53$.
In Fig. 4 it is possible to appreciate the dependence of colours with the age of galaxies, as it is expected from galaxy evolution. 
Thus, evolved galaxies are redder than the youngest ones. Nevertheless, the fraction of old galaxies is higher for bluer LINERs suggesting that these sources could have experienced some process ageing their stellar population.
For galaxies redder than $M_g-M_r \sim 0.9$ the fraction of galaxies older than the mean is $\approx$ 100\% and it is indistinguishable between LINER and control samples.
The errorbars in the figures were calculated by using bootstrap error resampling \citep{barrow84}.

The finding shows that LINER host galaxies belong to the most evolved galaxy population observed, independently of the local density environment. 
In this line, the recent works of \cite{sin13,belfiore16} suggest that the ubiquitous presence of this hot and evolved stellar population in extended LINERs strongly supports the assumption of post-AGB stars as the responsible mechanism to power the low-ionization emission lines.

\subsection{Morphology-density relation}

The relationship between density and morphology has been demonstrated by several authors. Thus, evolved elliptical galaxies are found inhabiting high density environments while blue disk galaxies commonly reside in low density regions \citep{Dress80,DML01}.
However, in the particular case of AGNs different studies \citep{PB06,coldwell09,padilla10,coldwell14} have found that these objects do not follow the expected morphology-density relation of galaxies without nuclear activity. 
Then, by exploring this relation for LINER host galaxies it could be possible to obtain clues about the existence of nuclear activity in the centre of the extended LINERs.

In this section we estimated the fraction of LINER and control galaxies older and redder than the average values ($ D_n (4000)> 1.53 $ and colours $ M_g-M_r> 0.76 $) as a function of their local density, $\Sigma_5$. In Fig.5 we show these fractions for both samples, it is observed than approximately 100\% of LINER galaxies are redder and older than the given mean values, in agree with the results of Fig. 3.
In addition, this result is not dependent on the local density, within the errorbars. Instead of that, the control sample shows a decreasing fraction of evolved galaxies at lower densities, as expected for the well known morphology-density relation.

It is important to notice that the mentioned differences between both sample could be directly related with the 
ionization mechanisms since control sample have been selected to match luminosity, concentration index, redshift and local 
density of LINERs. Hence, the colours and stellar age populations have demonstrated to be more sensitive to the involved physical processes by indicating different stages on galaxy evolution.  

\begin{figure}
\includegraphics[height=100mm,width=85mm]{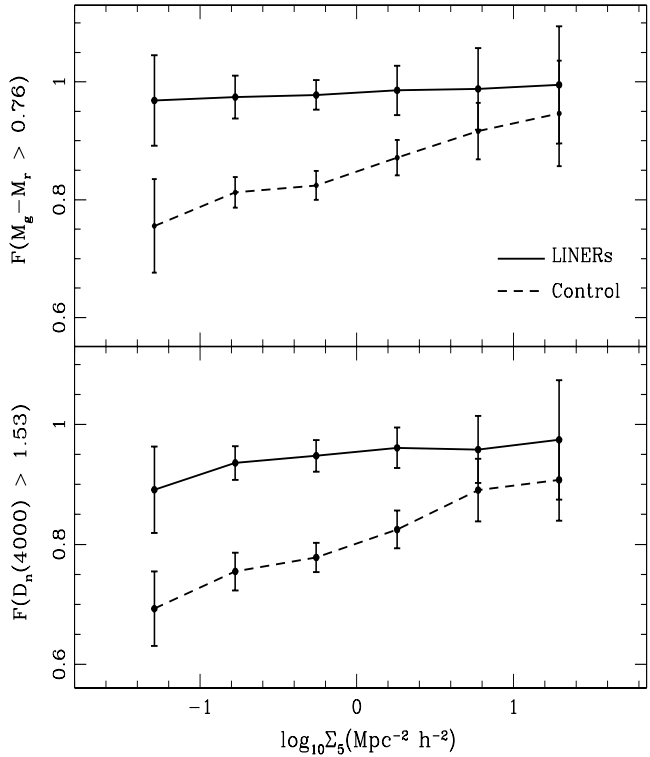}
\caption{Fraction of red ($M_g-M_r > 0.76$) and old ($D_n(4000)> 1.53$) galaxies
as function of the local density, $\Sigma_5$. The solid line corresponds to
the LINER sample and the dashed line to the control sample, respectively.}
\label{fig4}
\end{figure}

On the other hand, the stellar mass parameter correlates with colours, density and ages of galaxies. Thus, we expect lower values of stellar mass for blue and young typical galaxies (without special features as AGN). We examined the tendency 
of the fraction of red and old galaxies with respect to the stellar mass for both samples in three different ranges of density. We used the parameter $M^{\ast}$, in logarithmic scale, previously 
determined by \cite{kauff03b} where the method relies on spectral indicators relating to the stellar age population and the 
fraction of stars formed in recent bursts.

Figure 6 shows the fraction of red ($M_g-M_r > 0.76$) and old ($D_n(4000)> 1.53$) galaxies as function of the stellar mass, $M^\ast$ for three bins of local density given by $log_{10}\Sigma_5 < -0.5$, $-0.5 < log_{10}\Sigma_5 < 0.0$ and $log_{10}\Sigma_5 > 0.0$. These three ranges were selected to include approximately 33\% of the sample each one
and represent the low, intermediate and high density environment, respectively.
From this figure, it is possible to observe a clear tendency for galaxies in the control sample to become redder and older with increasing stellar mass. Moreover, these fractions are larger in the highest density range.
However, LINERs seem to be less sensitive to their host stellar mass. Their stellar age population shows a weak tendency to be younger for lower values of stellar mass, at low and intermediate densities, although the fraction of the $D_n(4000)$ parameter is almost  independent of the stellar mass at high densities. 

The noticeable effect observed for galaxy colour trends, in the upper panels of Fig.6,  where the fractions remain nearly 
constant for the whole range of $M^\ast$ (within the errorbars) independently of the local density environment, gives 
strong support to the lack of dependence between galaxy morphology and spatial density for LINER galaxies. 
These results present a good agreement with the work of \cite{coldwell17}. Furthermore,
the observed trends for LINER galaxies are thoroughly consistent with that found for AGN by
\cite{PB06,coldwell09,padilla10,coldwell14}.
In spite of the fact that extended LINERs have an old stellar population, consistent with post-AGB stars, which could be its main ionizing mechanism we can not discard the presence of AGN in the centre of these galaxies.

\begin{figure*}
\includegraphics[height=150mm,width=185mm]{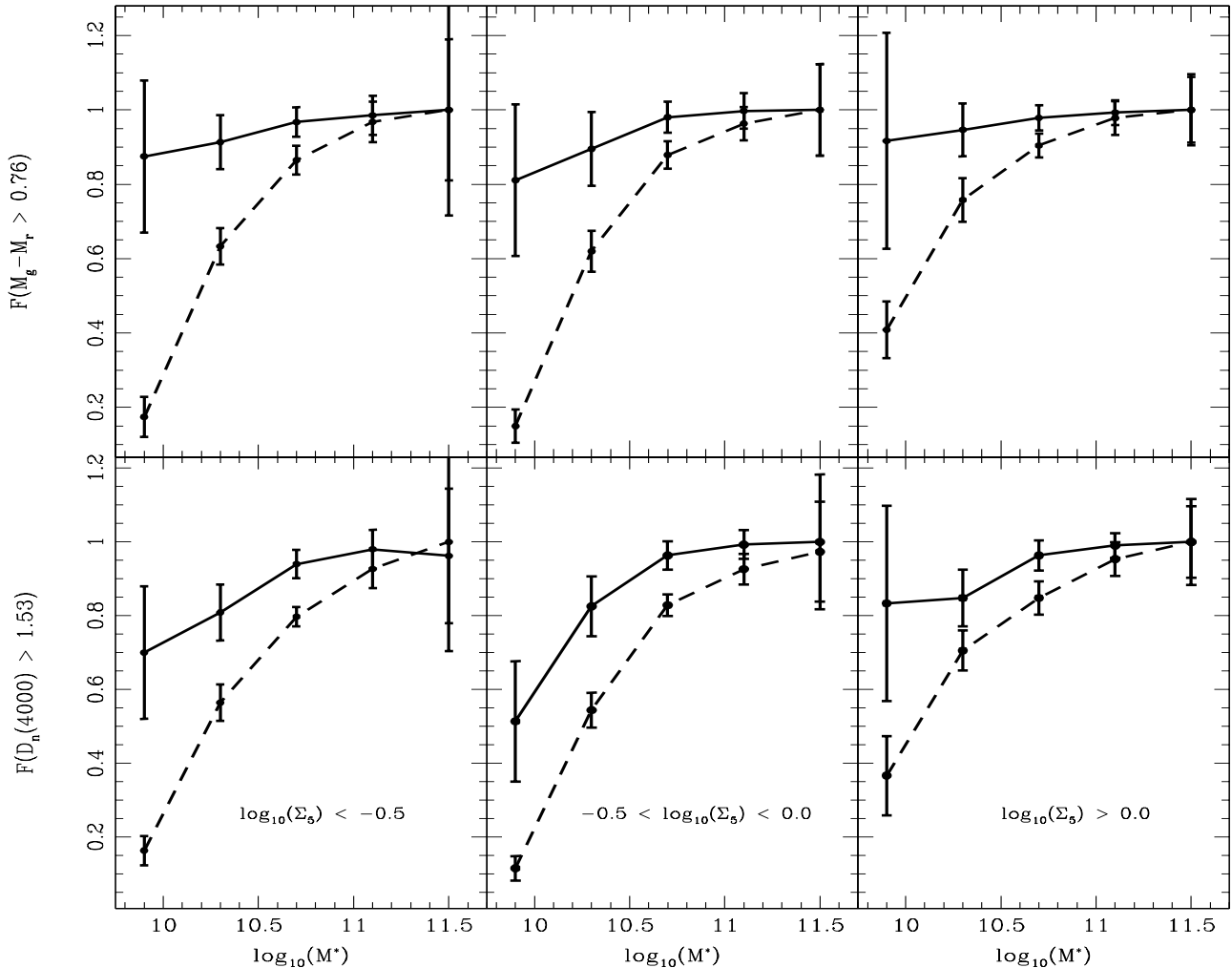}
\caption{Fraction of red ($M_g-M_r > 0.76$) and old ($D_n(4000)> 1.53$) galaxies
as function of the stellar mass, $M^\ast$, in logarithmic scale, for three different local density ranges. 
The solid line corresponds to the LINER sample and the dashed line to the control sample, respectively.}
\label{fig5}
\end{figure*}

\section{Discussion}

We derived a large LINER galaxy catalogue, by using SDSS-DR12 survey,  comprising 5560 objects in the redshift range $0.04 < z < 0.1$ . Taking into account the observational constrains and the median redshift of the galaxy sample, this is a  sample of extended LINERs.
In order to unveil the true environmental density dependence of the main  characteristics of galaxies hosting LINER and obtain clues about the mechanisms involved in the generation of low-ionization emission lines
we constructed a control sample of galaxies, without observed low ionization features, matched in redshift, luminosity, morphology and local density. This control sample have 5553 galaxies.

In this analysis we use a wide range of densities given by the local density estimator, $\Sigma_5$,  by exploring the dependence of this parameter with the stellar age populations and galaxy colours.
The distributions of $D_n(4000)$ and $M_g-M_r$ show that both LINER and control samples are older and redder than galaxies in the SDSS MGS in the redshift range considered in this work. Nevertheless, a significant difference is appreciated between colours and stellar age population of LINER and its respective control sample.
LINERs are noticeable older and redder than the control sample and this effect remains observable in the complete density range.  
In addition, an excess of old galaxies is detected for LINERs with respect to the control sample, at a given colour bluer than $M_g-M_r \approx 0.9$. This finding could suggest that LINERs have experienced some process that accelerate the evolution of their stellar population.

To study the dependence of the LINER host properties with the environmental density we calculated the fraction of red ($M_g-M_r > 0.76$) and old ($D_n(4000) > 1.53$) galaxies with respect to its local density. We find a remarkable difference between the behaviour of LINERs and control samples. While the control galaxies show an increasing fraction of red and old galaxies with the local density, approximately more than 90\% of the LINERs are redder and older than these mean values independently of their neighbourhood density.
This result implies that LINERs do not follow the known morphology-density relation shown for galaxies without low-ionization features.

In a more detailed analysis of the galaxy properties we estimated the fraction of red colours and old stellar population as a function of the stellar mass of galaxies ($log_{10}(M*)$) in three different ranges of local density. This specific variable strongly correlates with the processes involved in the galaxy evolution and density.  
The observed trends of the control sample clearly represent the expected morphology-density relation. Thus the fractions of both, colours and stellar age population, increase at larger values of stellar mass. 
In addition, these fractions are larger for the higher density ranges. Instead of that, the fractions of $D_n(4000)> 1.53$ for LINER galaxies are higher than that corresponding to the control sample.  Moreover, while a weak trend of increasing fractions can be observed at low and intermediate density ranges, this fraction does not depend on the host stellar mass in the range of highest density. Surprisingly, the fractions of $M_g-M_r$ LINER galaxy colours, redder than the mean value, is independent of the stellar mass and its remain constant, close to the 100\%,  for the three local density ranges.

Summarizing the main results we highlight the fact that LINER galaxies show a different behaviour than the control galaxy sample by analysing their colours and stellar age population. Certainly, LINERs have redder colours and older stellar populations independently of their stellar mass and the local density.
This finding is in agreement with the works of \cite{erac10,YanB12,sin13,belfiore16} suggesting a significant contribution from old and red hot post-asymptotic giant branch (post-AGB) stars as a driver of the photoionization. 
However, although the post-AGB hypothesis appears the most viable mechanism for extended LINERs,  these objects also exhibit features consistent with the presence of a central AGN, as it is the behaviour of the morphology-density relation.
The observational constraints make it difficult to draw definite conclusions. However
it is feasible that different mechanisms can coexist at different scales.

\section{Acknowledgments}
We would like to thank to the anonymous referee for the suggestions that helped to improve this paper.
G.V.C acknowledges the technical support from Dr. Rodrigo Romero and Estudio Conturso \& Asoc.
This work was supported in part by the Consejo Nacional de 
Investigaciones Cient\'ificas y T\'ecnicas de la Rep\'ublica Argentina 
(CONICET), the Consejo Nacional de Investigaciones Cient\'ificas, T\'ecnicas y de Creaci\'on Art\'istica de la 
Universidad Nacional de San Juan (CICITCA) and the Secretaría de Estado de Ciencia, Tecnolog\'ia e Innovaci\'on 
del Gobierno de San Juan (SECITI).

Funding for SDSS-III has been provided by the Alfred P. Sloan Foundation, the Participating Institutions, 
the National Science Foundation, and the U.S. Department of Energy Office of Science.
The SDSS-III web site is \emph{http://www.sdss3.org/}.
SDSS-III is managed by the Astrophysical Research Consortium for the Participating Institutions of the SDSS-III 
Collaboration including the University of Arizona, the Brazilian Participation Group, Brookhaven National Laboratory, 
Carnegie Mellon University, University of Florida, the French Participation Group, the German Participation Group, 
Harvard University, the Instituto de Astrofisica de Canarias, the Michigan State/Notre Dame/JINA Participation Group, 
Johns Hopkins University, Lawrence Berkeley National Laboratory, Max Planck Institute for Astrophysics, Max Planck 
Institute for Extraterrestrial Physics, New Mexico State University, New York University, Ohio State University, 
Pennsylvania State University, University of Portsmouth, Princeton University, the Spanish Participation Group, 
University of Tokyo, University of Utah, Vanderbilt University, University of Virginia, University of Washington, 
and Yale University.

{}

\label{lastpage}

\end{document}